\documentclass{ws-ijmpcs}

\begin{document}

\markboth{A. Austregesilo}
{Light-Meson Spectroscopy at GlueX}

%
\catchline{}{}{}{}{}
%

\title{Light-Meson Spectroscopy at GlueX}

\author{Alexander Austregesilo}

\address{Thomas Jefferson National Accelerator Facility\\
12000 Jefferson Avenue, Newport News, Virgina 23606, U.S.A.\\
aaustreg@jlab.org}

\author{(For the GlueX Collaboration)}


\maketitle

\begin{history}
\published{Day Month Year}
\end{history}

\begin{abstract}
GlueX at Jefferson Lab aims to study the light meson spectrum with an emphasis on the search for light hybrid mesons. To this end, a linearly-polarized $9\,$GeV photon beam impinges on a hydrogen target contained within a hermetic detector with near-complete neutral and charged particle coverage. In 2016, the experiment completed its commissioning and subsequently started to take data in its design configuration. With the size of the data set so far, GlueX already exceeds previous experiments for polarized photoproduction in this energy regime. A selection of early results will be presented, focusing on beam asymmetries for pseudo-scalar and vector mesons. The potential to make significant contributions to the field of light-meson spectroscopy is highlighted by the observation of several known meson resonances. Furthermore, the strategy to map the light meson spectrum with amplitude analysis tools will be outlined.

\end{abstract}

\section{Motivation}

The strong interaction is described by quantum chromodynamics (QCD) within the standard model of particle physics. The color-charged quarks interact by the means of gluons in analogy to quantum electrodynamics. However, the gluons interact among themselves due to their own color charge, which results in the confinement of the quarks into color-neutral objects. These hadrons are the relevant degrees of freedom at low energies, and they can be grouped into three-quark and quark-antiquark states, which are called baryons and mesons respectively. On the other hand, no known effect of the theory of strong interaction forbids other combinations of multiple quarks and excited gluons into color-neutral hadrons and numerical computations of QCD on the lattice even predict the manifestation of these so-called exotic states in the spectrum of hadrons\cite{Dud13}. The question of the existence of exotic hadrons is one of the unsolved problems in modern particle physics. 

In the constituent quark model, mesons are characterized by a set of quantum numbers $J^{PC}$. However, some combinations of these quantum numbers are forbidden for a quark-antiquark state. These so-called spin-exotic states are the smoking gun in the search for physics beyond the constituent quark model, since they cannot mix with conventional mesons. It is therefore important to study not only the cross section but the angular distribution in order to determine the spin of the states and disentangle them from the other broad and overlapping resonances within the light quark meson spectrum.

\section{Experimental Setup}

The GlueX experiment at Jefferson Lab\footnote{Thomas Jefferson National Accelerator Facility} is part of a global effort to study the spectrum of hadrons, with the polarized photon beam contributing a unique feature. Polarization of the photons is achieved by coherent Bremsstrahlung of the primary electron beam of up to 12\,GeV on a thin diamond radiator. The scattered electrons are used to tag the energy of the photon beam. With a collimator suppressing the incoherent Bremsstrahlung spectrum, a linear polarization of 40\% is achieved in the coherent peak at 9\,GeV. In order to cancel apparatus effects, the polarization plane is rotated by steps of 45$^\circ$ into four different orientations during physics data taking. The degree of polarization is measured using the phenomenon of triplet photoproduction\cite{Dug17}. 

\begin{figure}[pt]
\centerline{\includegraphics[width=.5\textwidth]{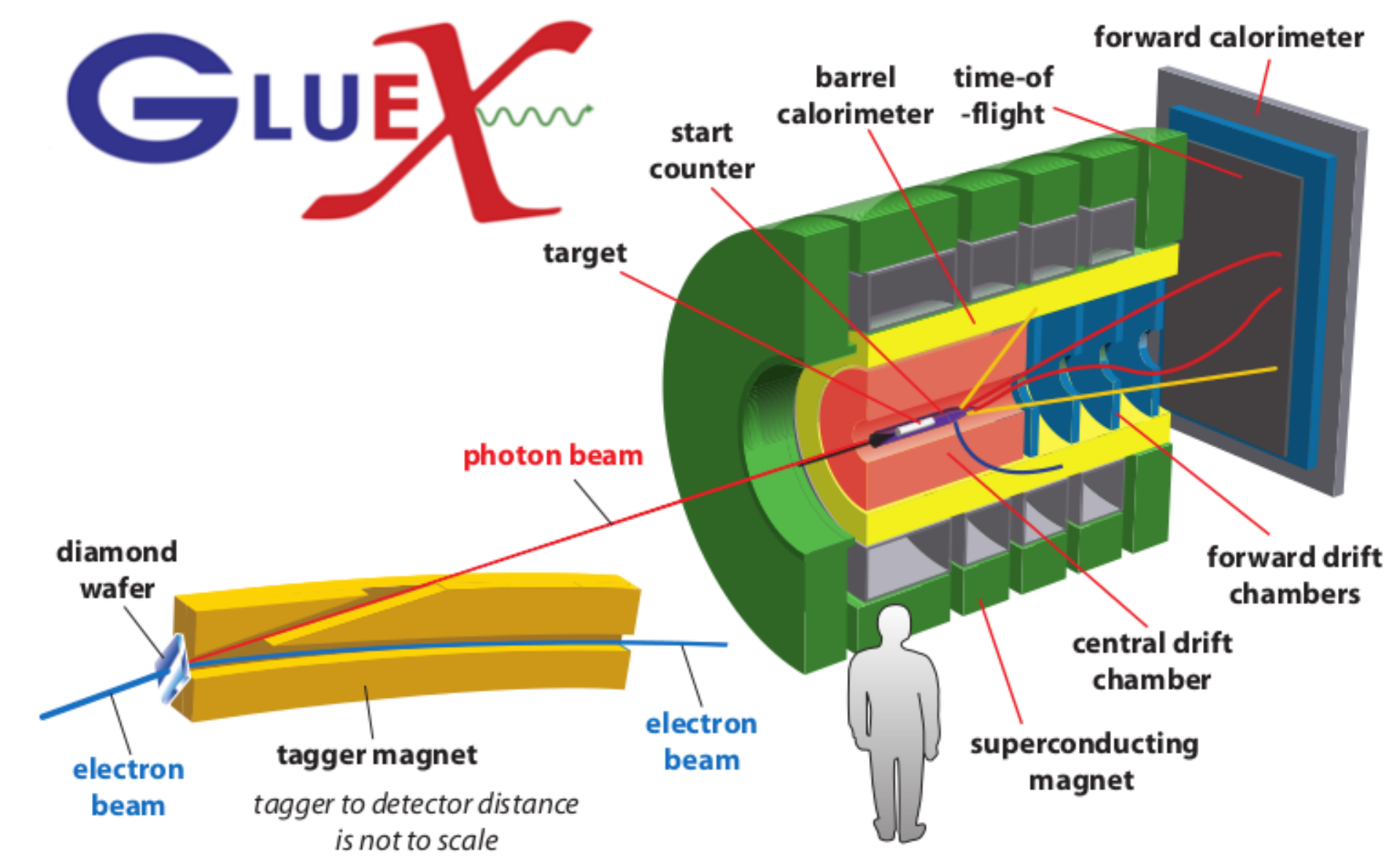}}
\vspace*{8pt}
\caption{The GlueX experimental setup.\label{fig:det}}
\end{figure}

The beam impinges on a liquid hydrogen target. Assuming vector meson dominance, a wide variety of states are accessible. The beam intensity in the coherent peak of up to $5 \cdot 10^7$ photons per second provides a sufficiently high reaction rate to study even rare processes. The GlueX detector was designed to map the light quark meson spectrum up to masses approximately 3\,GeV$/c^2$ with full acceptance for all decay modes. A superconducting solenoid magnet with a 2\,T field houses the target, central and forward drift chambers, and a barrel calorimeter\cite{Bea18} (see~Fig.~\ref{fig:det}). A forward calorimeter completes the forward photon acceptance and a time-of-flight counter provides particle identification capability.

The beamline and the detector were fully commissioned in spring 2016, when an initial physics data set was recorded. It effectively amounts to about 80 hours of running and most of the results presented in this document were obtained with this data set. In spring 2017, the first phase of GlueX was started and will continue through 2018. About 20\% of the full projected data set using the baseline detector were already collected to this day. From 2019 onwards, an upgraded detector will continue taking data with an even higher luminosity.

\section{Early Results}

For the analysis and interpretation of the data, it is indispensable to understand the underlying photoproduction mechanism. To this end, an early priority of the GlueX collaboration is the study of polarization transfers and spin-density matrix elements (SDMEs) of the production processes. Once these are sufficiently understood, we will measure the cross sections of know mesons before we will be confident enough to identify exotic hadrons by the means of amplitude analysis. A dialogue with the authors of theoretical models will be very beneficial for these steps as the analysis of high statistical precision data requires revisiting previously used simplifying assumptions. With this in mind, the GlueX experiment is collaborating with JPAC\footnote{Joint Physics Analysis Center}\cite{} in order to establish capable analysis frameworks and robust theoretical models.

\subsection{Beam Asymmetries}

A thorough understanding of meson photoproduction in the GlueX energy regime is necessary for the interpretation of the data. The azimuthal dependence of the polarized production cross section can be parametrized as $\sigma_\text{pol} (\phi) = \sigma_\text{unpol} \left [ 1 - P_\gamma \Sigma \cos2\phi \right ]$, where $P_\gamma$ is the measured beam polarization and $\phi$ the azimuthal angle of the produced meson with respect to the polarization plane of the photon beam. The beam asymmetry $\Sigma$ is extracted from the data as well as its dependence on the squared four-momentum transfer $t$. This quantity is sensitive to the quantum numbers of the Regge exchange. Systematic effects are canceled by rotating the polarization plane by $90^\circ$.

Using the initial 2016 data set, we have published the beam asymmetry for the production of the neutral mesons $\pi^0$ and $\eta$. A detailed discussion of these results can be found in Ref.~\refcite{Glu17,Zha18}. The results exceed the only existing measurement for $\pi^0$ in precision and are the first ones for $\eta$ at these energies. The measurements are compared to various models and may constrain the background to baryon resonance production. The fact that $\Sigma$ is close to unity indicates the dominance of vector exchanges\cite{Mat15}. These single pseudoscalar production measurements are being extended to multiple decays of the $\eta$ and $\eta'$ as well as to charged mesons $\pi^\pm$ and $K^\pm$ with the much larger 2017 data set. Charge exchange reactions are particularly interesting as they were used in previous analyses to study disputed hybrid meson candidates\cite{CLA09}.

\begin{figure}[pt]
\centerline{\includegraphics[width=.45\textwidth]{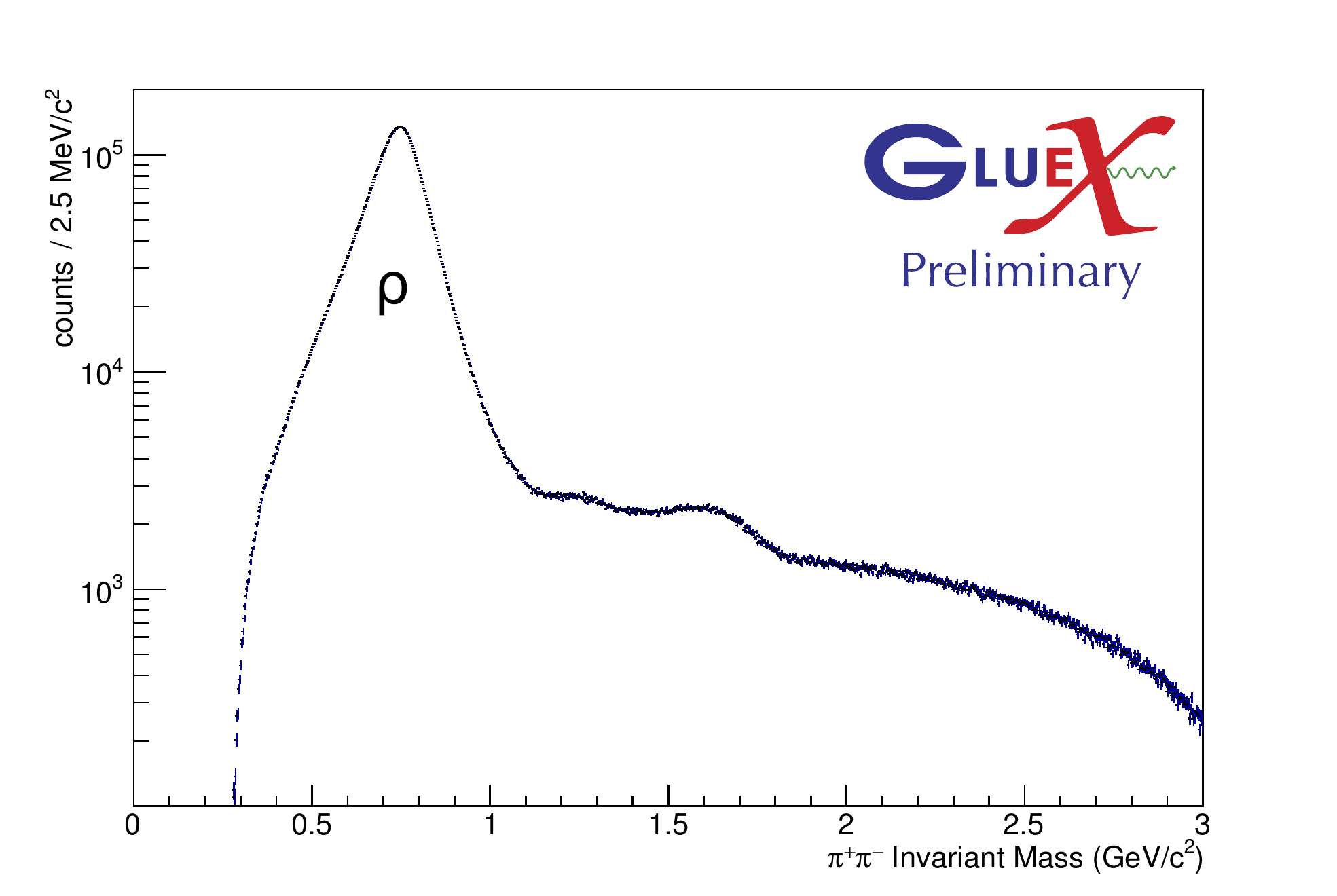}\includegraphics[width=.45\textwidth]{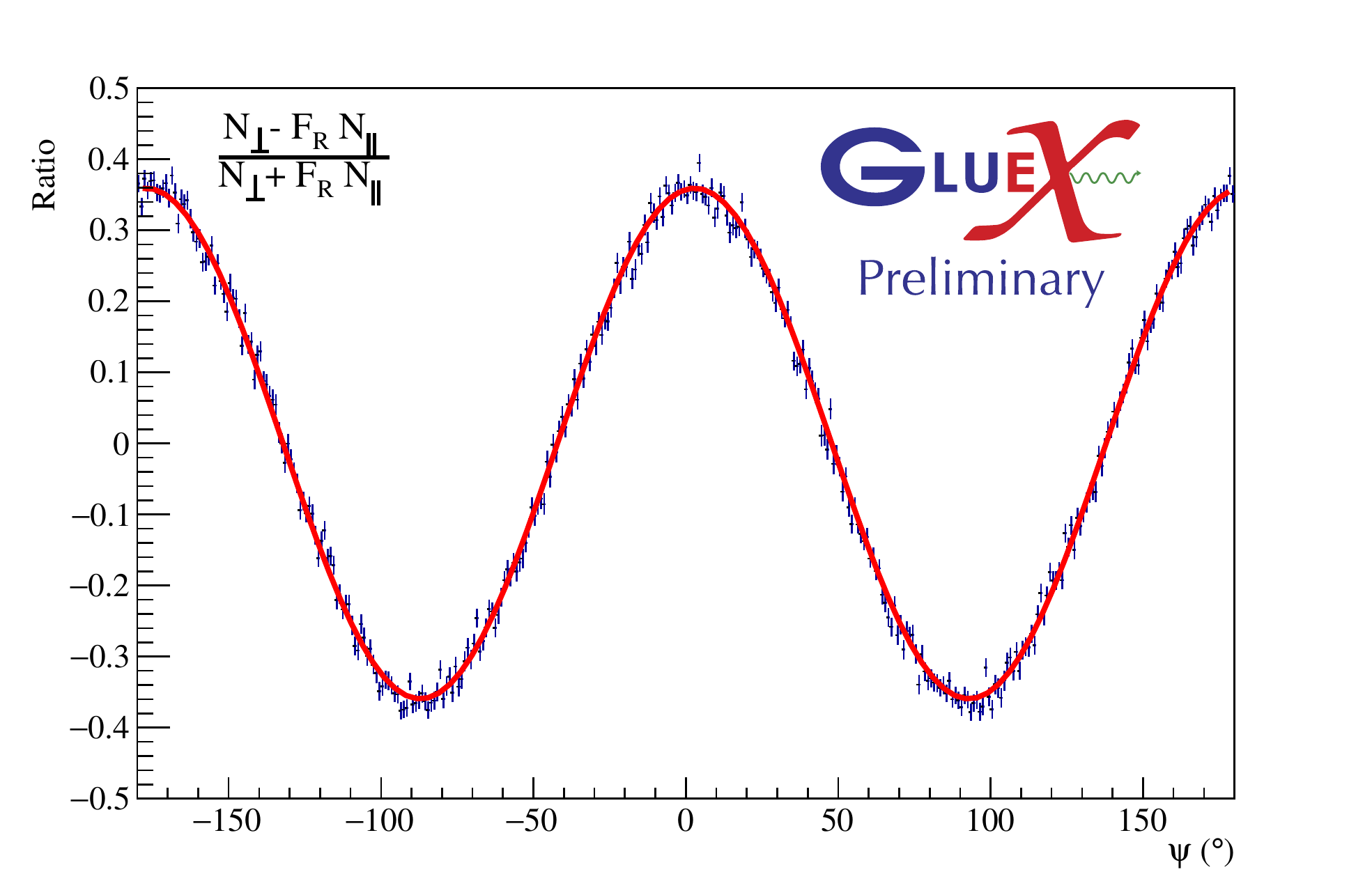}}
\vspace*{8pt}
\caption{(Left) Invariant mass of the $\pi^+\pi^-$ system. (Right) Decay asymmetry of $\rho$(770) meson.\label{fig:rho}}
\end{figure}

SDMEs generalize this concept for the production of mesons with spin. We are currently studying the production of $\rho$, $\omega$, and $\phi$ mesons via the angular distributions of their decays.
Fig.~\ref{fig:rho} (right) shows the asymmetry observed in $\rho$(770) decays for illustration of the anticipated precision.

\subsection{Prospects for Spectroscopy}

Multi-particle final states can not only be used to study the production of ground state mesons. The extent of the data set already shows signs for excited states in the spectra, that surpass the previously observed number of events by orders of magnitude. One impressive example is shown in Fig.~\ref{fig:rho} (left), where the invariant mass of the exclusively produced $\pi^+\pi^-$ system is strongly dominated by the production of the $\rho$(770) meson. However, a clear enhancement can be discerned near the mass of $1.6\,$GeV$/c^2$, which is consistent with previous observations of an excited $\rho$ state\cite{Abe84}. In addition, the production of the tensor state $f_2$(1270) may play a role as well.

The GlueX detector has an almost hermetic electromagnetic calorimetry coverage which makes it ideal to study decays into multiple photons. Exclusive events with a reconstructed recoil proton and four photons in the final state select very clean samples of $\pi^0\pi^0$ and $\pi^0\eta$ production. Clear evidence for the scalar meson $a_0$ and the tensor mesons $f_2$ and $a_2$ can be discerned in the invariant mass spectra, respectively\cite{Zha18}.

\begin{figure}[pt]
\centerline{\includegraphics[width=.5\textwidth]{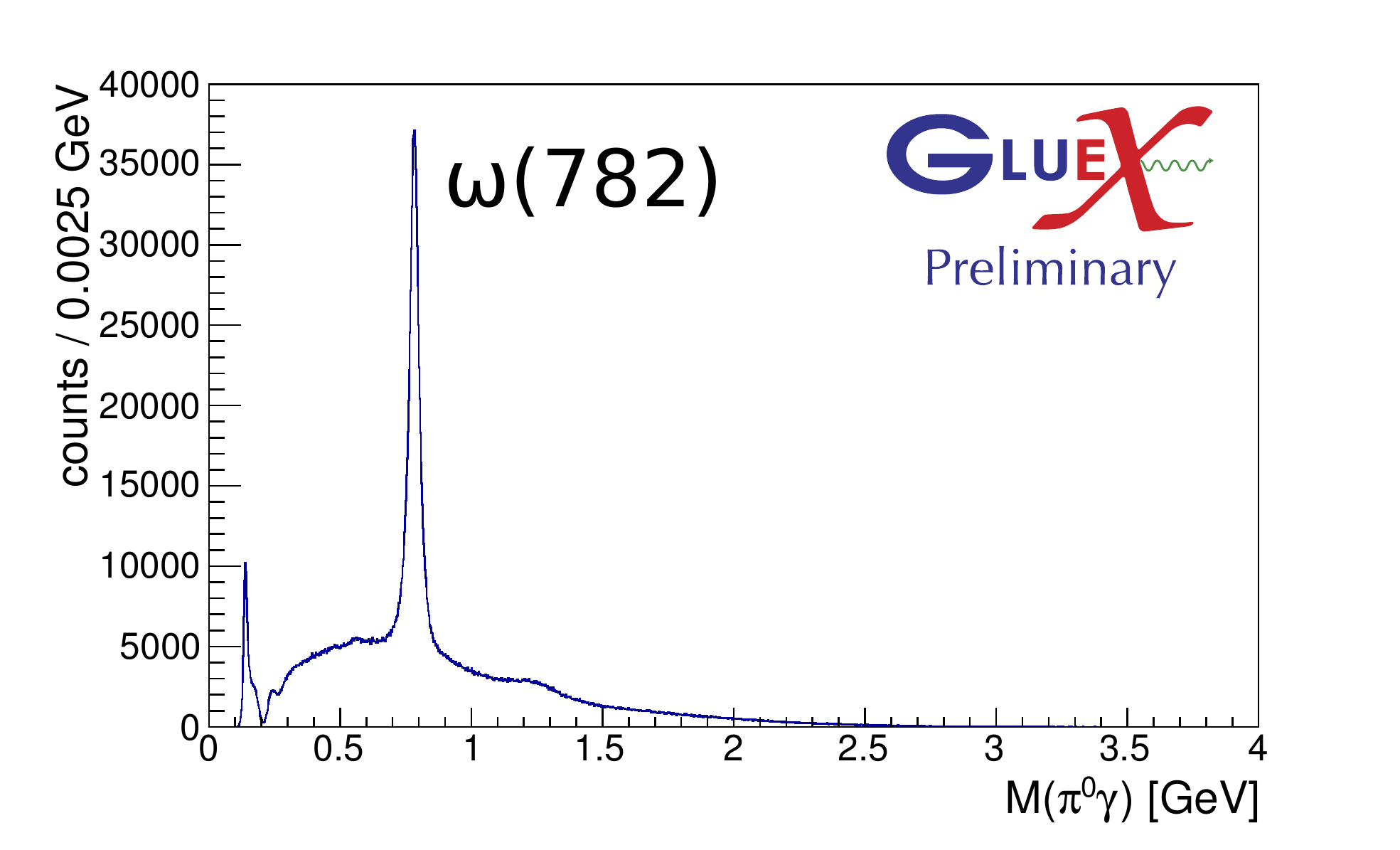}\includegraphics[width=.5\textwidth]{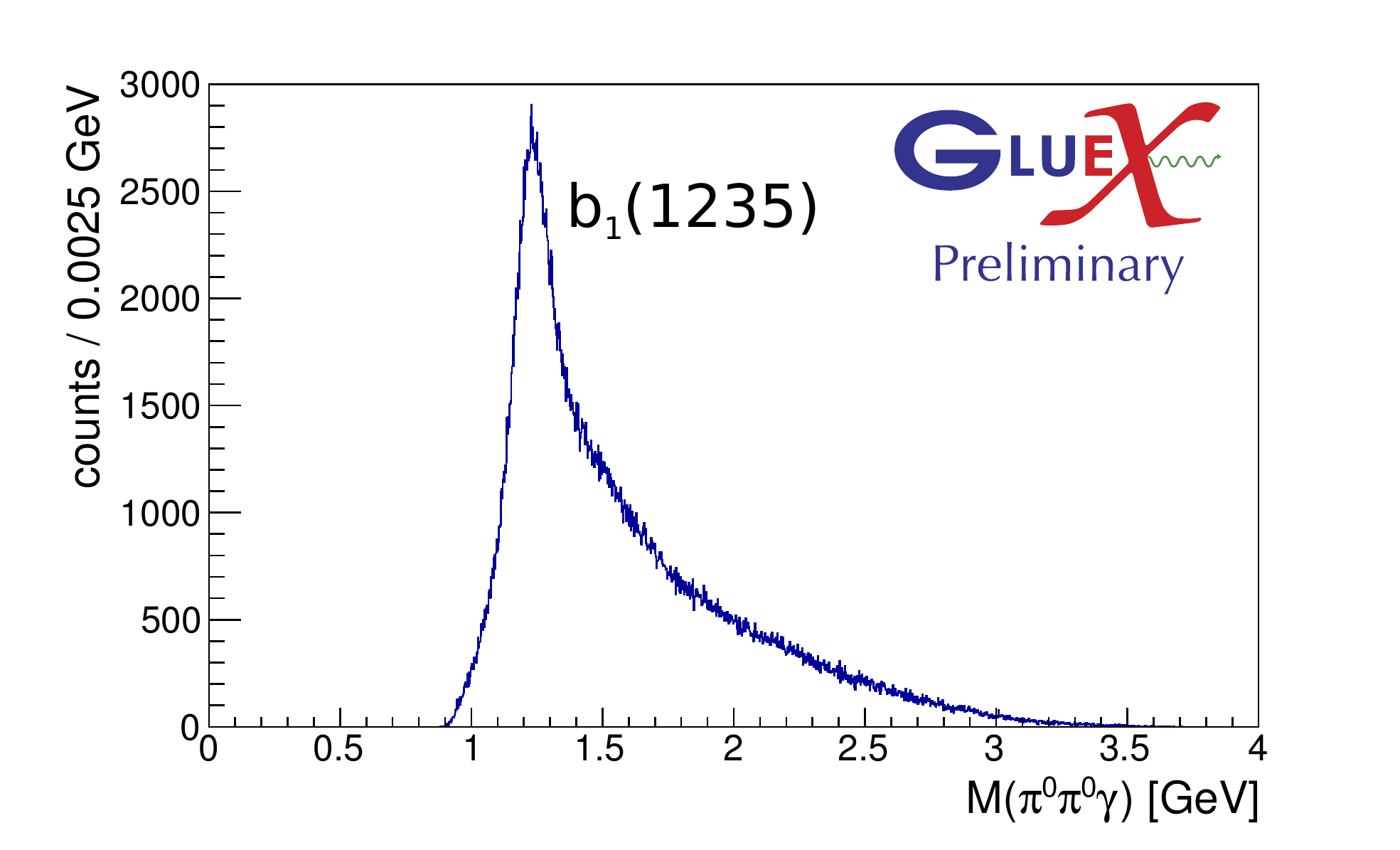}}
\vspace*{8pt}
\caption{Invariant mass spectra of the reaction $\gamma +p \rightarrow 5\gamma +p$: (left) $\pi^0\gamma$ (right) $\omega\pi^0$.	 ~~~~~~~~~~~~~~~~~~~~~~~~~~~~ Not background subtracted.\label{fig:b1}}
\end{figure}

The reaction $\gamma + p \rightarrow 5\gamma + p$ is dominated by the decay of $\omega$(782) into $\pi^0\gamma$. Restricting the invariant mass of the $3\gamma$ subsystem to the region of the omega, one can nicely select the decay of $b_1$(1235) into $\omega\pi^0$ (see~Fig.~\ref{fig:b1}). This meson is particularly interesting, as it has been reported as a decay product of the spin-exotic signals $\pi_1$(1600) and $\pi_1$(2000)\cite{Bnl05}.

Even with six photons in the final state, we are able to select clean exclusive samples of $\eta$ decaying into $3\pi^0$ and $\eta'$ decaying into $\eta\pi^0\pi^0$. This demonstrates that the GlueX detector meets the requirements for an excellent spectroscopy program.

\subsection{$J/\psi$ Photoproduction}

Apart from the main focus on light meson spectroscopy, the recorded data yield a range of opportunistic measurements. One particularly interesting example is the measurement of the cross section of $J/\psi$ photoproduction near threshold. This very clean laboratory for charmonium production not only provides information about the gluonic structure of the nucleus\cite{Bro01}, but allows us to directly study the photocouplings of the pentaquark states states observed at LHCb\cite{LHCb15,Hil16}. Only very little information is available in the threshold region which translates into photon beam energies below $12\,$GeV.

\begin{figure}[pt]
\centerline{\includegraphics[width=.59\textwidth]{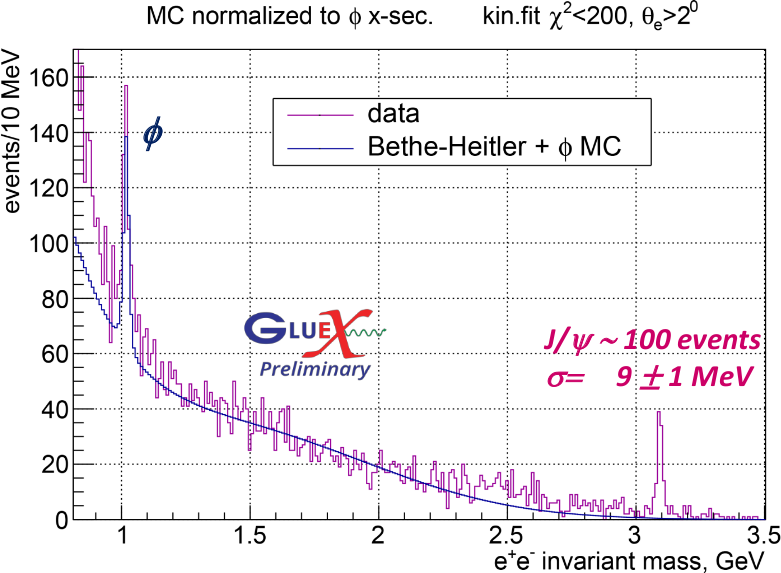}\hfill\includegraphics[width=.35\textwidth]{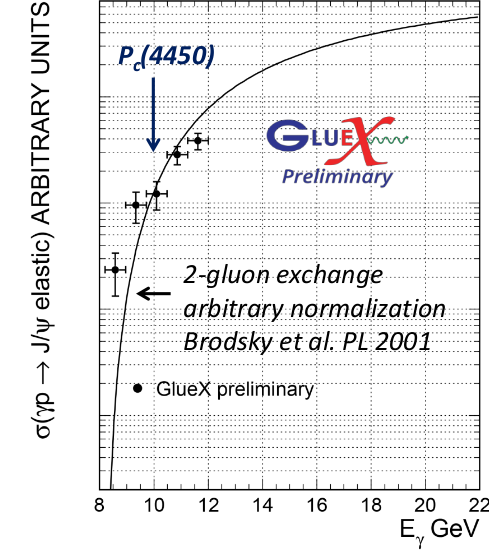}}
\vspace*{8pt}
\caption{(Left) Invariant mass of $e^+e^-$ pair. (Right) $J/\psi$ yield as a function of beam energy.\label{fig:jpsidata}}
\end{figure}

With the initial GlueX data, we were able to reconstruct about 100 events of $J/\psi$ decaying into $e^+e^-$ pairs, where the electrons are identified by their electromagnetic showers in the calorimeters. 
Fig.~\ref{fig:jpsidata} (left) shows the di-electron mass spectrum, where a clean peak can be discerned near $3.1\,$GeV/$c^2$ and the background is well described by a simulation of the Bethe-Heitler process. This constitutes the first observation of charmonium at Jefferson Lab and will allow us to measure the cross section as well as the $t$-slope of the production process.

Fig.~\ref{fig:jpsidata} (right) shows preliminary results for the arbitrarily normalized event rate as a function of the incident beam energy to illustrate the statistical precision of the data. The final GlueX data set is expected to provide up to an order of magnitude more events than shown here.

\section{Conclusion and Outlook}

The GlueX detector has been successfully commissioned and has started its physics data taking campaign in Spring 2017. Even though these data represent only approximately 20\% of the full data set, the analysis of a variety of physics topics is already ongoing. The photoproduction mechanism for single pseudoscalar and vector mesons is studied as a first step towards amplitude analysis of the light meson spectrum. The comparison with previous measurements and theoretical model helps us to understand the detector acceptance and the systematic uncertainties.

The observation of many well-known meson resonances as well as the first successful reconstruction of the $J/\psi$ at Jefferson Lab demonstrates good prospects for a broad physics program with initial GlueX data. The mapping of the entire light meson spectrum will be possible, with a precise measurement of the properties of known resonances and, ultimately, the candidates for exotic mesons.

The completion of the initial GlueX running is expected in 2019, after which several upgrades are planned. A DIRC detector will be added in the forward region to improve $\pi/K$ separation\cite{Ste16}. This and a factor of five or more higher luminosity will enhance the program to strange hadron spectroscopy.

\section*{Acknowledgments}

This material is based upon work supported by the U.S. Department of Energy, Office of Science, Office of Nuclear Physics under contract DE-AC05-06OR23177.


\end{document}